\documentclass[12pt]{article}
\usepackage{epsfig}
\usepackage{amsfonts}
\usepackage{amscd}
\usepackage{latexsym}
\usepackage{amsmath,amssymb}
\usepackage{verbatim}
\usepackage{setspace}
\usepackage{color}
\usepackage{cite}

\usepackage[textheight=9in, textwidth=6.5in, letterpaper]{geometry}

\usepackage{hyperref}
\numberwithin{equation}{section}

   \makeatletter
  \let\over=\@@over \let\overwithdelims=\@@overwithdelims
  \let\atop=\@@atop \let\atopwithdelims=\@@atopwithdelims
  \let\above=\@@above \let\abovewithdelims=\@@abovewithdelims
\renewcommand\section{\@startsection {section}{1}{\z@}%
                                   {-3.5ex \@plus -1ex \@minus -.2ex}
                                   {2.3ex \@plus.2ex}%
                                   {\normalfont\large\bfseries}}

\renewcommand\subsection{\@startsection{subsection}{2}{\z@}%
                                     {-3.25ex\@plus -1ex \@minus -.2ex}%
                                     {1.5ex \@plus .2ex}%
                                     {\normalfont\bfseries}}

\begin{document}
\unitlength = 1mm

\ \\
\vskip 1cm
\begin{center}

{\textbf{\large{More on the conformal mapping of quasi-local masses: The Hawking-Hayward case}}}

\vspace{0.8cm}
Fay\c{c}al Hammad\footnote{fhammad@ubishops.ca}

\vspace{1cm}

{\it  Physics Department \& STAR Research Cluster, Bishop's University\\
2600 College Street, Sherbrooke, (QC) J1M 1Z7, Canada\\
Physics Department, Champlain College-Lennoxville\\
2580 College Street, Sherbrooke, (QC) J1M 0C8, Canada}
\begin{abstract}
The conformal transformation of the Hawking-Hayward quasi-local mass is reexamined. It has been found recently that the conformal transformation of the latter exhibits the 'wrong' conformal factor compared to the way usual masses transform under conformal transformations of spacetime. We show, in analogy with what was found recently for the Misner-Sharp mass, that unlike the purely geometric definition of the Hawking-Hayward mass, the latter exhibits the 'right' conformal factor whenever expressed in terms of its material content via the field equations. The case of conformally invariant scalar-tensor theories of gravity is also examined. The equivalence between the Misner-Sharp mass and the Hawking-Hayward mass for spherically symmetric spacetimes manifests itself by giving identical peculiar behaviors under conformal transformations.
\end{abstract}

\vspace{1.0cm}

\end{center}


\pagestyle{empty}
\pagestyle{plain}

\pagenumbering{arabic}

\section{Introduction}\label{sec:1}
The concept of quasi-local energy is very important for the study of many aspects of the physics of spacetime. Its application ranges from the study of black hole physics and spacetime thermodynamics, and all the way to cosmology. The need to define the gravitational energy in a given region of spacetime has led many authors to introduce various definitions of the concept. The most well-known concept is certainly the Misner-Sharp mass \cite{MS}, applicable for spherically symmetric spacetimes. The other very important definition of quasi-local mass is the Hawking-Hayward mass \cite{Haw,Hay}, as the latter could be used for more general configurations of spacetime. More importantly, the Hawking-Hayward mass reduces to the Misner-Sharp mass in the special case of a spherically symmetric spacetime \cite{Hay}. The Hawking-Hayward quasi-local mass has been successfully applied for another derivation of the first law for black holes \cite{Mukohyama}. The Misner-Sharp quasi-local mass has been successfully applied both to the study of black hole thermodynamics \cite{Hayward, Cai1} and to rederive the Friedmann equations in cosmology \cite{Cai2, Cai3, Gong, Akbar}.

The other very important concept in the study of spacetime and gravitational physics is the concept of conformal transformations. Conformal transformations constitute a great tool in the investigation of gravitational theories since one goes, via these transformations, from one frame to another where the equations become either more familiar or just easier to manipulate. Conformal transformations are a special category of spacetime mappings that change the metric by a conformal factor, and hence affect the lengths and the proper time of a given physical system. The physical consequences of a conformal transformation is that a length $\ell$ gets transformed into a new length $\tilde{\ell}=\Omega\ell$, the proper time of a freely falling observer $\tau$ changes into a new one given by $\tilde{\tau}=\Omega\tau$, and a mass/energy $m$ changes into a new mass/energy $\tilde{m}$ related to the old one by $\tilde{m}=\Omega^{-1}m$. The latter transformation could actually be guessed on dimensional grounds since mass/energy has the dimensions of an inverse length in the natural units of $\hbar=c=1$.

The combination of both concepts, the Misner-Sharp and Hawking-Hayward quasi-local masses on one hand and conformal transformations on the other, has been the subject of recent investigations in Ref.~\cite{FaraoniVitagliano,Prain}. The result was that, counterintuitively, these quasi-local masses do not transform as usual masses do as one might have expected. Indeed, what was found in Ref.~\cite{FaraoniVitagliano} is that the Misner-Sharp mass $m_{_{MS}}(t,r)$ transforms into a new mass $\tilde{m}_{_{MS}}(t,r)=\Omega[m_{_{MS}}(t,r)+...]$ and in Ref.~\cite{Prain} it was found that the Hawking-Hayward mass $m_{_{HH}}(t,r)$ transforms into the new mass $\tilde{m}_{_{HH}}(t,r)=\Omega[m_{_{HH}}(t,r)+...]$, where the ellipsis represent in general complicated geometrical quantities in both cases. Thus, these quasi-local masses do not transform the way real masses do but rather they transform the way a length does.

A moment of thought, however, reveals that this result could hardly be surprising since what one finds actually in geometric terms is not really the geometric equivalent of the mass $m(t,r)$, but the geometric equivalent of the product $Gm(t,r)$, where $G$ is Newton's gravitational constant, which indeed has the dimensions of a length and, as such, could not transform otherwise than as a length. In fact, as we shall see it in detail in this paper, this observation is well supported by the fact that the Hawking-Hayward mass recovers the normal conformal transformation of a usual mass within conformally invariant scalar-tensor theories, such as Brans-Dicke theory \cite{BD}. Indeed, in such theories, Newton's gravitational constant $G$ is replaced by the effective gravitational 'constant' $G_{e\mathrm{ff}}=\phi^{-1}$, where $\phi$ is the Brans-Dicke field. Thereby, quasi-local mass concepts give the geometric equivalent of $\phi^{-1}m$. In other words, the material mass in this case is the geometric mass times $\phi$. Since after a conformal transformation the scalar $\phi$ becomes $\Omega^{-2}\phi$, what we end up with in the conformal frame is simply the geometric mass, which transforms like a length, times $\Omega^{-2}$; and the 'right' factor $\Omega^{-1}$ is thus automatically recovered.

Now this does not raise any issue as long as one keeps to what one has started with. In other words, if one starts with a purely geometric entity one obtains at the end the correct conformal transformation of that geometric entity. If, on the other hand, one starts with a material entity one will find at the end the correct conformal transformation of that entity. The issue rises only when one attempts to \textit{deduce} the conformal transformation of either entities after having taken the other \textit{different} entity as a starting point. The obstacle that stands in the face of such an approach is the conformal non-invariance of Einstein's field equations which relate the source, the matter on the right-hand side, to the effect, the geometry on the left-hand side. Trying to find the conformal transformation of the source by first computing the conformal transformation of the effect can only give unexpected results.

Schematically, this translates into the following statement. Start from the field equations $Geometry=\kappa\,Matter$ in the original frame (Jordan frame). Now, suppose that after a conformal transformation $Geometry$ transforms into $\widetilde{Geometry}$ and matter transforms into $\widetilde{Matter}$ in the new frame (Einstein frame). Then, after a conformal transformation one does not end up with field equations of the form $\widetilde{Geometry}=\kappa\,\widetilde{Matter}$, but rather with equations of the form $\widetilde{Geometry}=\kappa\,Matter+More\,Geometry$, where $More\,Geometry$ stands for a bunch of additional geometric terms that, as we will see in detail in the next section, could be interpreted as an induced energy content. With this simplified scheme, we can see clearly where the issue rises. Searching for the conformal transformation of $Geometry$ is fine since it gives the geometric equivalent of $\kappa\,Matter$ in the Einstein frame. However, if one attempts to deduce the conformally transformed $\widetilde{Matter}$ by identifying it with $\widetilde{Geometry}/\kappa$ one necessarily finds an incorrect result as the field equations in the Einstein frame do not assume the form $\widetilde{Geometry}=\kappa\,\widetilde{Matter}$.

In the next section we shall see how to make this simplified scheme more precise by examining the familiar Hawking-Hayward quasi-local mass. Since the conformal transformation of the geometric definition of the latter has been treated in great detail in Ref.~\cite{Prain}, we shall only recall here the main results \footnote{For completeness, we shall add in the present paper the conformal transformation of the geometric Hawking-Hayward mass within scalar-tensor theories, a subject which has not been touched upon in Ref.~\cite{Prain}. The reason is that the result will help us make here a precise comparison between the conformal transformations of Misner-Sharp and Hawking-Hayward masses, and it will shed more light on all the subtleties of the behavior of these masses under conformal mappings.}, as our main focus will be on the use of the field equations to perform the transformation. The issue raised by the Hawking-Hayward quasi-local mass is more subtle as the latter concept was not born from material considerations in contrast to the Misner-Sharp mass which was motivated by the physics of gravitational collapse. The issue raised by the Hawking-Hayward mass is that it gives a wrong conformal factor only if one abstains from using the field equations inside the integral defining the mass. The right conformal factor, as required by dimensional analysis, is recovered as soon as one appeals to the field equations.

In summary, then, the importance of both the concept of quasi-local masses and the conformal mappings of spacetime calls for a detailed study of the intricacies of both concepts. The aim of the present paper is to expose the different facets of the conformal transformation of the Hawking-Hayward quasi-local mass in order to contribute to our understanding of the curious way the quasi-local energies transform during a conformal mapping of spacetime, and hence increase our understanding of the latter.

The outline of the paper is as follows. In Sec.~\ref{sec:2}, we recall the geometric definition of the Hawking-Hayward mass. We also recall the main results found in Ref.~\cite{FaraoniCQG} concerning the use of the field equations to define the Hawking-Hayward mass. In Sec.~\ref{sec:3} we investigate all the facets of the conformal transformation of the mass and show how the result depends on whether or not one relies on the field equations to define the mass. In Sec.~\ref{sec:5}, we discuss the similarities between the Misner-Sharp and the Hawking-Hayward masses. We end this paper with a conclusion section in which we summarize our main findings and address the main conclusions.

\section{The Hawking-Hayward mass from the field equations}\label{sec:2}
The original idea behind the Hawking quasi-local mass \cite{Haw} was to integrate the dynamically independent parts of the gravitational field over a spatial compact and orientable 2-surface $\mathcal{S}$ bounding the three-dimensional region one is interested in finding its corresponding gravitational energy. The formula of Hawking for quasi-local energy was then of the form $M_{_{H}}\sim\sqrt{A}\int_{\mathcal{S}}\mu(R^{h}+\theta_{+}\theta_{-})$, where $\mu$ is the area 2-form on the 2-surface $\mathcal{S}$ and $A$ is the total area of the latter. $R^{h}$ is the Ricci scalar corresponding to the induced metric $h_{\mu\nu}$ on the 2-surface, and $\theta_{+}=\frac{1}{2}h^{\mu\nu}\mathcal{L}_{u-r}h_{\mu\nu}$ and $\theta_{-}=\frac{1}{2}h^{\mu\nu}\mathcal{L}_{v-s}h_{\mu\nu}$ measure the expansion of the 2-surface with respect to the two evolution vectors $u^{\mu}$ and $v^{\mu}$, respectively, which set the double-null foliation of spacetime \cite{Hay}. The symbol $\mathcal{L}$ represents the usual Lie-derivative. The shift vectors on the 2-surface are $r^{\mu}$ and $s^{\mu}$, obtained by projecting the evolution vectors on the 2-surface: $r^{\mu}=h^{\mu}_{\nu}u^{\nu}$ and $s^{\mu}=h^{\mu}_{\nu}v^{\nu}$.

Although the Hawking definition yields the correct zero mass when computed over a sphere in a flat spacetime, it gives a wrong result for general 2-surfaces in flat spacetime. The Hayward extension \cite{Hay} of the Hawking mass was introduced with the goal to remedy this situation. This extended Hawking quasi-local mass, also called Hawking-Hayward quasi-local mass, always gives the right answer for any generic spatial 2-surface bounding a 3-volume in space. The exact expression of such a mass reads \cite{Hay},
\begin{align}\label{HHDefiniton}
M_{_{HH}}&=\frac{1}{8\pi G}\sqrt{\frac{A}{16\pi}}\int_{\mathcal{S}}\mu\left(R^{(h)}+\theta_{+}\theta_{-}-\frac{1}{2}\sigma_{\mu\nu}^{+}\sigma^{\mu\nu}_{-}
-2\omega_{\mu}\omega^{\mu}\right).
\end{align}
Here, $G$ is Newton's gravitational constant and indices are lowered or raised using the full metric $g_{\mu\nu}$ of spacetime. The tensors $\sigma^{+}_{\mu\nu}=(h^{\rho}_{\mu}h^{\lambda}_{\nu}-\frac{1}{2}h_{\mu\nu}h^{\rho\lambda})\mathcal{L}_{u-r}h_{\rho\lambda}$ and $\sigma^{-}_{\mu\nu}=(h^{\rho}_{\mu}h^{\lambda}_{\nu}-\frac{1}{2}h_{\mu\nu}h^{\rho\lambda})\mathcal{L}_{v-s}h_{\rho\lambda}$ are called the shear tensors corresponding to the evolution of the 2-surface. The vector $\omega_{\mu}=e^{m}h_{\mu\nu}(\mathcal{L}_{u}s^{\nu}-\mathcal{L}_{v}r^{\nu}-\mathcal{L}_{r}s^{\nu})$ is called the twist vector and quantifies the spatial twist, in the evolution of the 2-surface, with respect to its two evolution vectors $u^{\mu}$ and $v^{\mu}$. The variable $m$ is called the scaling function and gives the scalar product of the two null normals to the foliation \cite{Hay}. It turns out that the integrand in the above expression could be simplified by expressing it in terms of the most often used four-dimensional Riemann tensor since $R^{(h)}+\theta_{+}\theta_{-}-\frac{1}{2}\sigma_{\mu\nu}^{+}\sigma^{\mu\nu}_{-}={h}^{\mu\rho}h^{\nu\lambda}R_{\mu\nu\rho\lambda}$ \cite{Hay}. Furthermore, in order to simplify also our subsequent formulas, we introduce here the reduced Hawking-Hayward mass, $m_{_{HH}}=32\pi^{\frac{3}{2}}M_{_{HH}}$, and we set Newton's constant $G$ equal to unity. Then, we have the much simpler formula,
\begin{equation}\label{HHReducedDefinition}
m_{_{HH}}=\sqrt{A}\int_{\mathcal{S}}\mu\left(h^{\mu\rho}h^{\nu\sigma}R_{\mu\nu\rho\sigma}-2\omega_{\mu}\omega^{\mu}\right).
\end{equation}
This is the purely geometric definition of the Hawking-Hayward quasi-local mass that we shall work with. It must be noted here that the integrand in (\ref{HHReducedDefinition}) is actually the Hamiltonian 2-form referred to the compact orientable 2-surface $\mathcal{S}$\cite{Hay}. This remark will be useful to us later when we suggest a physical interpretation of the new terms that appear after conformally transforming the mass.

On the other hand, when dealing with scalar-tensor theories of gravity, like the Brans-Dicke theory which we shall examine in detail below, the inverse $G^{-1}$ of Newton's gravitational constant is replaced by the spacetime-dependent scalar field $\phi(x)$. Thereby, the effective gravitational 'constant' should contribute in the form $G_{\mathrm{eff}}^{-1}=\phi$ inside the integral of (\ref{HHReducedDefinition}) as the 'constant' varies from one point to the other on the compact 2-surface $\mathcal{S}$. In this case, the geometric definition (\ref{HHReducedDefinition}) should then be replaced by,
\begin{equation}\label{HHReducedDefinitionST}
m_{_{HH}}=\sqrt{A}\int_{\mathcal{S}}\mu\phi\left(h^{\mu\rho}h^{\nu\sigma}R_{\mu\nu\rho\sigma}-2\omega_{\mu}\omega^{\mu}\right).
\end{equation}

Now, it is possible to express this mass in terms of the material content by making use of the field equations. Indeed, after recalling the first contracted Gauss relation, $R_{\mu\nu\rho\sigma}=C_{\mu\nu\rho\sigma}+g_{\mu[\rho}R_{\sigma]\nu}-g_{\nu[\rho}R_{\sigma]\mu}-\frac{1}{3}g_{\mu[\rho}g_{\sigma]\nu}R$, one is able to substitute for the Ricci tensor $R_{\mu\nu}$ its value in terms of the energy-momentum tensor as dictated by the field equations. Doing so in the case of general relativity, for which one has the 'simple' Einstein equations $R_{\mu\nu}-\frac{1}{2}g_{\mu\nu}R=8\pi T_{\mu\nu}$, turns expression (\ref{HHReducedDefinition}) into \cite{FaraoniCQG},
\begin{equation}\label{HHinGR}
m_{_{HH}}=\sqrt{A}\int_{\mathcal{S}}\mu\left[h^{\mu\rho}h^{\nu\sigma}C_{\mu\nu\rho\sigma}-2\omega_{\mu}\omega^{\mu}+8\pi\left(h^{\mu\nu}T_{\mu\nu}
-\frac{2}{3}T\right)\right].
\end{equation}
The same procedure could be used in extended theories of gravity, like scalar-tensor theories, albeit the results would be more complicated. Indeed, when substituting the value of the Riemann tensor in terms of the energy-momentum tensor as dictated by the field equations of the simplest scalar-tensor theories of gravity with a single scalar field $\phi$, namely, the Brans-Dicke theory whose gravitational Lagrangian is $(16\pi)^{-1}\sqrt{-g}(\phi R-\phi^{-1}\omega(\phi)\partial_{\mu}\phi\partial^{\mu}\phi)-V(\phi)$, expression (\ref{HHReducedDefinition}) takes on the following more complicated form \cite{FaraoniCQG}:
\begin{multline}\label{HHinST}
m_{_{HH}}=\sqrt{A}\int_{\mathcal{S}}\mu\phi\Bigg[h^{\mu\rho}h^{\nu\sigma}C_{\mu\nu\rho\sigma}-2\omega_{\mu}\omega^{\mu}+\frac{8\pi}{\phi}\left(h^{\mu\nu}T_{\mu\nu}
-\frac{2T}{3}\right)
\\+\frac{h^{\mu\nu}\nabla_{\mu}\nabla_{\nu}\phi}{\phi}+\frac{\omega}{\phi^{2}}\left(h^{\mu\nu}\nabla_{\mu}\phi\nabla_{\nu}\phi
-\frac{1}{3}\nabla^{\rho}\phi\nabla_{\rho}\phi\right)+\frac{V}{3\phi}\Bigg].
\end{multline}
Here, $T_{\mu\nu}$ and $T$ are, respectively, the energy-momentum tensor and its trace of ordinary matter obtained from the Lagrangian $L_{matter}$. Notice that in both theories a separation into a contribution from matter and another from pure geometry is possible. We shall do that here in view of a our later reference to it. Thus, expression (\ref{HHinST}), as well as expression, (\ref{HHinGR}) could be split into a purely material part $m_{_{HH}}^{M}$ and a purely geometric part $m_{_{HH}}^{G}$ as follows: $m_{_{HH}}=m_{_{HH}}^{M}+m^{G}_{_{HH}}$, where,
\begin{multline}\label{HHinSTSplitting}
m_{_{HH}}^{M}=\sqrt{A}\int_{\mathcal{S}}8\pi\mu\left(h^{\mu\nu}T_{\mu\nu}
-\frac{2T}{3}\right),
\\m_{_{HH}}^{G}=\sqrt{A}\int_{\mathcal{S}}\mu\phi\Bigg[h^{\mu\rho}h^{\nu\sigma}C_{\mu\nu\rho\sigma}-2\omega_{\mu}\omega^{\mu}
+\frac{h^{\mu\nu}\nabla_{\mu}\nabla_{\nu}\phi}{\phi}+\frac{\omega}{\phi^{2}}\left(h^{\mu\nu}\nabla_{\mu}\phi\nabla_{\nu}\phi
-\frac{1}{3}\nabla^{\rho}\phi\nabla_{\rho}\phi\right)+\frac{V}{3\phi}\Bigg].\\\!\!
\end{multline}

In the case of general relativity the field $\phi$ in the second line of (\ref{HHinSTSplitting}) should be set equal to a constant and the pure gravitational part reduces to $m_{_{HH}}^{G}=\sqrt{A}\int_{\mathcal{S}}\mu(h^{\mu\rho}h^{\nu\sigma}C_{\mu\nu\rho\sigma}-2\omega_{\mu}\omega^{\mu})$.
Note that this splitting of the mass into a purely geometric part and a purely material part is done according to the usual interpretation of the Brans-Dicke scalar field $\phi$ as being a geometric entity, rather than a material field, whose energy-momentum tensor is hence not included inside the $T_{\mu\nu}$ of matter but kept in the form of a sum of explicit derivatives of $\phi$ and its potential $V$.
But it must be kept in mind, though, that such a splitting is only possible when using the field equations to define the Hawking-Hayward mass. In the next section we shall see all the difference the use of the field equations brings to the conformal transformation of the Hawking-Hayward mass.
\section{Conformal mapping of the geometric Hawking-Hayward mass}\label{sec:3}
Let us now examine the consequences the use the field equations to define the Hawking-Hayward mass has on the conformal transformation of the latter. A general conformal mapping of spacetime has the effect of transforming the metric $g_{\mu\nu}(x)$ into the new metric $\tilde{g}_{\mu\nu}(x)$ such that,
\begin{equation}\label{ConformalDefinition}
\tilde{g}_{\mu\nu}=\Omega^{2}(x)g_{\mu\nu},
\end{equation}
where $\Omega(x)$, called the conformal factor, is a spacetime-dependent smooth and non-vanishing function. In order to find the form the Hawking-Hayward mass takes in the Einstein frame one uses the following results \cite{Prain} on the conformal transformation of the various entities appearing inside the integral (\ref{HHReducedDefinition})\footnote{It should be noted that the conformal transformation of $\omega^{\mu}$ adopted here is not really unique as the vector $\omega^{\mu}$ depends on the vectors $r^{\mu}$ and $s^{\mu}$, and the scalar $m$, which express only the coordinate freedom on the surfaces $\mathcal{S}$ of the foliation. This does not, however, restrict our arguments as one uses the Hawking-Hayward mass only after having chosen a specific foliation. See the discussions in \cite{Prain,FaraoniCQG} for more details.}:
\begin{align}\label{ConfTansformationsA}
&\qquad\qquad\tilde{\mu}=\Omega^{2}\mu,\qquad\tilde{A}=\int_{\mathcal{S}}\mu\Omega^{2},\qquad\tilde{h}^{\mu\nu}=\Omega^{-2}h^{\mu\nu},\qquad \tilde{\omega}_{\mu}\tilde{\omega}^{\mu}=\Omega^{-2}\omega_{\mu}\omega^{\mu},\nonumber
\\&\tilde{R}_{\mu\nu\rho\sigma}=\Omega^{2}R_{\mu\nu\rho\sigma}
+\Big[2\Omega g_{\sigma[\mu}\nabla_{\nu]}\nabla_{\rho}\Omega+4g_{\rho[\mu}\nabla_{\nu]}\Omega\nabla_{\sigma}\Omega
+g_{\sigma\mu}g_{\nu\rho}\nabla_{\lambda}\Omega\nabla^{\lambda}\Omega-(\rho\leftrightarrow\sigma)\Big].
\end{align}
Substituting these inside (\ref{HHReducedDefinition}), one easily finds then, within general relativity, the following transformation of the Hawking-Hayward mass when adopting its definition based on purely geometric entities \cite{Prain}:
\begin{equation}\label{ConfGeoHH}
\tilde{m}_{_{HH}}=\sqrt{\frac{\tilde{A}}{A}}\left(m_{_{HH}}+m_{_{HH}}^{I}\right),
\end{equation}
where the notation,
\begin{equation}\label{mHHInduced}
m_{_{HH}}^{I}=2\sqrt{A}\int_{\mathcal{S}}\mu
\left[h^{\mu\nu}\left(\frac{2\nabla_{\mu}\Omega\nabla_{\nu}\Omega}{\Omega^{2}}-\frac{\nabla_{\mu}\nabla_{\nu}\Omega}{\Omega}\right)
-\frac{\nabla^{\rho}\Omega\nabla_{\rho}\Omega}{\Omega^{2}}\right],
\end{equation}
has been introduced here to denote the conformally 'induced' geometric Hawking-Hayward mass. The integrand on the right-hand side of (\ref{mHHInduced}) might be interpreted, in the light of the remark made below the definition (\ref{HHReducedDefinition}), as being the projection, on the 2-surface $\mathcal{S}$, of the Hamiltonian associated with the induced energy momentum-tensor by the 'work' done during the conformal deformation $\Omega(x)$ of spacetime. Since this integral will be frequently appearing below, introducing such a notation in this paper will make things more transparent and easier to track.

The most important remark we should include here about the above result, though, is the fact that after a conformal transformation the Hawking-Hayward mass scales like a length does and not like real masses do. This can be seen more explicitly by dividing both sides of the result (\ref{ConfGeoHH}) by $\sqrt{\tilde{A}}$. Equation (\ref{ConfGeoHH}) then just says that the mass 'per unit length' $m_{_{HH}}/\sqrt{A}$ is invariant under a conformal transformation, modulo of course additional terms representing 'energy per unit' length coming from the conformal geometry. It is, however, obvious that unless the Hawking-Hayward mass $m_{_{HH}}$ scales like a length, i.e. like $\sqrt{A}$ does, the ratio $m_{_{HH}}/\sqrt{A}$ can never be a conformal invariant. We will come back to this fact below, which is behind the issue raised by conformal transformations of the Hawking-Hayward mass.

In view of the comparison with what happens in scalar-tensor theories that we shall make below, we shall give here the result as it applies for the simple case of spherical symmetry, for which the area $A$ transforms as $\tilde{A}=\Omega^{2}A$ and the conformal factor depends only on the radial coordinate $r$. The result (\ref{ConfGeoHH}) simplifies in this case to,
\begin{equation}\label{ConfGeoHHSphere}
\tilde{m}_{_{HH}}=\Omega\left(m_{_{HH}}+m_{_{HH}}^{I}\right).
\end{equation}
Thus, in analogy with what happens with the Misner-Sharp mass based on its purely geometric definition \cite{FaraoniVitagliano, Hammad}, the formula clearly displays the same pattern: an overall factor $\Omega$ multiplies both the original mass and the geometrically induced one, in contrast to what happens for usual masses during a conformal transformation.

On the other hand, within the framework of scalar-tensor theories of gravity, namely the Brans-Dicke theory to which we restrain ourselves in this paper, the geometric definition one should adopt as we saw above is (\ref{HHReducedDefinitionST}). Therefore, after a conformal transformation we won't find the simple pattern (\ref{ConfGeoHH}), but we find instead the following transformed mass:
\begin{multline}\label{ConfGeoHHST}
\tilde{m}_{_{HH}}=\sqrt{\tilde{A}}\int_{\mathcal{S}}\frac{\mu\phi}{\Omega^{2}}\left(h^{\mu\rho}h^{\nu\sigma}R_{\mu\nu\rho\sigma}-2\omega_{\mu}\omega^{\mu}\right)
\\+2\sqrt{\tilde{A}}\int_{\mathcal{S}}\frac{\mu\phi}{\Omega^{2}}
\left[h^{\mu\nu}\left(\frac{2\nabla_{\mu}\Omega\nabla_{\nu}\Omega}{\Omega^{2}}-\frac{\nabla_{\mu}\nabla_{\nu}\Omega}{\Omega}\right)
-\frac{\nabla^{\rho}\Omega\nabla_{\rho}\Omega}{\Omega^{2}}\right].
\end{multline}
In deriving this result, we have used, beside the conformal transformations (\ref{ConfTansformationsA}), the fact, which we will explain below, that the field $\phi$ transforms as $\tilde{\phi}=\Omega^{-2}\phi$.

The result (\ref{ConfGeoHHST}) is completely different from the result (\ref{ConfGeoHH}). This can be seen more clearly within the simple case of spherical symmetry. In this case, we can take out $\Omega^{-2}$ from the two integrals in (\ref{ConfGeoHHST}) and we can use $\tilde{A}=\Omega^{2}A$. We find,
\begin{equation}\label{ConfGeoHHSTSphere}
\tilde{m}_{_{HH}}=\Omega^{-1}\left(m_{_{HH}}+m^{I}_{_{HH}}\right),
\end{equation}
where $m_{_{HH}}$ is the initial mass within the scalar-tensor theory, given by (\ref{HHReducedDefinitionST}), and the geometrically induced mass $m_{_{HH}}^{I}$ is here given by,
\begin{equation}\label{mHHInducedST}
m_{_{HH}}^{I}=2\sqrt{A}\int_{\mathcal{S}}\mu\phi
\left[h^{\mu\nu}\left(\frac{2\nabla_{\mu}\Omega\nabla_{\nu}\Omega}{\Omega^{2}}-\frac{\nabla_{\mu}\nabla_{\nu}\Omega}{\Omega}\right)
-\frac{\nabla^{\rho}\Omega\nabla_{\rho}\Omega}{\Omega^{2}}\right].
\end{equation}
The result (\ref{ConfGeoHHSTSphere}) could easily be confronted with the result (\ref{ConfGeoHHSphere}) to see the great difference conformally invariant scalar-tensor theories make on the conformal transformation of the Hawking-Hayward geometric mass. We clearly see that, in contrast to what one gets from general relativity, the mass transforms just as normal masses do under conformal transformations. This result illustrates well the point we made intuitively in the Introduction.

Now the results (\ref{ConfGeoHH}) and (\ref{ConfGeoHHST}) are unique and apply for general relativity and its Brans-Dicke extension, respectively. As soon as one uses the field equations corresponding to the specific theory one chooses, however, the final results one finds differ not only between the different theories, which is totally natural, but even within the same theory. We first examine below the case of general relativity and then we proceed to the more complicated case of scalar-tensor theories.
\section{Field equations-based conformal mapping of Hawking-Hayward mass}\label{sec:4}
\subsection{Within general relativity}
In order to illustrate the above claim, let us start from expression (\ref{HHinGR}) obtained from the field equations within the framework of general relativity. To go from that expression to its conformally transformed counterpart, one only needs to use, in addition to the transformations (\ref{ConfTansformationsA}), the following results \cite{Dabrowski} on the remaining terms in (\ref{HHinGR}):
\begin{equation}\label{ConfTansformationsB}
\tilde{C}_{\mu\nu\rho\sigma}=\Omega^{2}C_{\mu\nu\rho\sigma}, \qquad\tilde{T}_{\mu\nu}=\Omega^{-2}T_{\mu\nu},\qquad\tilde{T}=\Omega^{-4}T.
\end{equation}
The substitution of these inside integral (\ref{HHinGR}) gives straightforwardly the following result for the conformally transformed Hawking-Hayward mass in general relativity:
\begin{equation}\label{ConfHHinGR}
\tilde{m}_{_{HH}}=\sqrt{\tilde{A}}\int_{\mathcal{S}}\mu\frac{8\pi}{\Omega^{2}}\left(h^{\mu\nu}T_{\mu\nu}
-\frac{2}{3}T\right)+\sqrt{\frac{\tilde{A}}{A}}m^{G}_{_{HH}}.
\end{equation}
Here, $m^{G}_{_{HH}}$ is the geometric part of the Hawking-Hayward mass, defined below Eq.~(\ref{HHinSTSplitting}) for general relativity, and corresponds to the pure gravitational energy of the vacuum region of spacetime, i.e. when the latter is devoid of any material content. It is obtained from (\ref{HHinGR}) by setting the energy-momentum tensor of matter $T_{\mu\nu}=0$ there.

The first thing to notice about the result (\ref{ConfHHinGR}) is that it is built from two parts which did not acquire the same conformal factor. Indeed, by dividing both sides of the identity by $\sqrt{\tilde{A}}$, we see that the mass 'per unit length' $\tilde{m}_{_{HH}}/\sqrt{\tilde{A}}$ is not invariant anymore because, unlike the second term $m_{_{HH}}^{G}/\sqrt{A}$, the first term on the right-hand side of the equation displays the form of 'a mass per unit length'$/\Omega^{2}$. This behavior, however, could only appear if the term in question transforms under conformal transformations like normal masses do.

The above observation becomes even clearer when one restricts to the special case of spherical symmetry, for which one then uses a conformal factor $\Omega(r)$ that depends only on the radial coordinate $r$. In such a situation, the factor $\Omega(r)$ can be taken out of the integral. Moreover, in that case we also have a simple conformal transformation for the area: $\tilde{A}=\Omega^{2}A$. The result (\ref{ConfHHinGR}) will then simply read,
\begin{equation}\label{ConfHHinGRSphere}
\tilde{m}_{_{HH}}=\Omega^{-1}m_{_{HH}}^{M}+\Omega\,m^{G}_{_{HH}}.
\end{equation}
From this, the resulting pattern followed during a conformal transformation by the different parts is now clearly recognizable. The material part and the geometric part behave differently.

Now this fact can actually be easily understood as follows. When imposing the field equations on the geometric entities appearing inside the Hawking-Hayward integral (\ref{HHReducedDefinition}), after a conformal transformation the latter becomes split into two parts, just as it occurs in Einstein field equations themselves which become: $\tilde{G}_{\mu\nu}=\kappa T_{\mu\nu}+T_{\mu\nu}^{\Omega}$; the first term being the material content and the second is the geometrically induced content. What happens then is that after a conformal transformation the different entities get filtered out and each part behaves its own way, as it is displayed in the result (\ref{ConfHHinGR}), due to their different natures. Notice, though, that what appeared inside the Hawking-Hayward mass is not exactly the Einstein tensor, $G_{\mu\nu}=R_{\mu\nu}-\frac{1}{2}g_{\mu\nu}R$, but rather the combination $R_{\mu\nu}-\frac{1}{6}g_{\mu\nu}R$. Indeed, we have $h^{\mu\rho}h^{\nu\sigma}R_{\mu\nu\rho\sigma}=h^{\mu\rho}h^{\nu\sigma}C_{\mu\nu\rho\sigma}+h^{\mu\nu}(R_{\mu\nu}-\frac{1}{6}g_{\mu\nu}R)$. We will come back to this remark below.

\subsection{Within scalar-tensor theories}
Let us now examine the case of the simplest scalar-tensor theories of gravity by applying the conformal transformation on integral (\ref{HHinST}) obtained from the field equations of the theory. To do that, one needs, in addition to the transformations (\ref{ConfTansformationsA}) and (\ref{ConfTansformationsB}), the following transformations of the remaining terms in (\ref{HHinST})\cite{Dabrowski, FaraoniBook}:
\begin{equation}\label{ConfTansformationsC}
\tilde{\phi}=\Omega^{-2}\phi,\qquad\omega\rightarrow\tilde{\omega}.
\end{equation}
Here, the scalar field $\phi$ transforms as an inverse length squared for dimensional reasons, for in Brans-Dicke theory this field represents the inverse $G_{eff}^{-1}$ of the effective Newton's gravitational 'constant' and is multiplied by the Ricci scalar inside the action \cite{FaraoniBook, Fujii, De felice, Capozziello, Nojiri}. For other scalar-tensor theories of gravity the field $\phi$ behaves differently and the invariance conditions would be different but the arguments developed here remain valid. On the other hand, the conformal transformation of the Brans-Dicke parameter $\omega(\phi)$ is left unspecified here as it depends on the theory. It must be recalled, though, that the transformation of $\omega(\phi)$ is what constrains conformal invariance of the Brans-Dicke scalar-tensor theory of gravity. As we will see below, just by demanding that the induced mass has the same form as the one found from the purely geometric definition (\ref{HHReducedDefinition}), it is sufficient to recover the same constraint on the transformation of $\omega(\phi)$ that makes the field equations and the action invariant.

Substituting all three groups of transformations (\ref{ConfTansformationsA}), (\ref{ConfTansformationsB}), and (\ref{ConfTansformationsC}) when writing the conformally transformed integral (\ref{HHinST}) in the Einstein frame, i.e. when putting tildes on the different terms, one easily finds the following result for the conformal transformation of the Hawking-Hayward mass in Brans-Dicke type scalar-tensor theories of gravity:
\begin{align}\label{ConfHHinST}
&\tilde{m}_{_{HH}}=\sqrt{\tilde{A}}\int_{\mathcal{S}}\frac{\mu\phi}{\Omega^{2}}\Bigg[h^{\mu\rho}h^{\nu\sigma}C_{\mu\nu\rho\sigma}-2\omega_{\mu}\omega^{\mu}+\frac{8\pi}{\phi}\left(h^{\mu\nu}T_{\mu\nu}
-\frac{2T}{3}\right)\nonumber
\\&\qquad\qquad\qquad+\frac{h^{\mu\nu}\nabla_{\mu}\nabla_{\nu}\phi}{\phi}+\frac{\omega}{\phi^{2}}\left(h^{\mu\nu}\nabla_{\mu}\phi\nabla_{\nu}\phi
-\frac{1}{3}\nabla^{\rho}\phi\nabla_{\rho}\phi\right)+\frac{V}{3\phi}\Bigg]\nonumber
\\&+\sqrt{\tilde{A}}\int_{\mathcal{S}}\frac{\mu\phi}{\Omega^{2}}h^{\mu\nu}\Bigg[\frac{10\nabla_{\mu}\Omega\nabla_{\nu}\Omega}{\Omega^{2}}
-\frac{2\nabla_{\mu}\nabla_{\nu}\Omega}{\Omega}-\frac{6\nabla_{\mu}\phi\nabla_{\nu}\Omega}{\Omega\phi}
-\frac{2g_{\mu\nu}\nabla_{\rho}\Omega\nabla^{\rho}\Omega}{\Omega^{2}}+\frac{g_{\mu\nu}\nabla_{\rho}\phi\nabla^{\rho}\Omega}{\Omega\phi}\Bigg]\nonumber
\\&+\sqrt{\tilde{A}}\int_{\mathcal{S}}\frac{\mu\phi}{\Omega^{2}}\Bigg[\frac{4\tilde{\omega}}{\Omega^{2}}
\left(h^{\mu\nu}\nabla_{\mu}\Omega\nabla_{\nu}\Omega-\frac{1}{3}\nabla_{\rho}\Omega\nabla^{\rho}\Omega\right)-\frac{4\tilde{\omega}}{\Omega\phi}
\left(h^{\mu\nu}\nabla_{\mu}\Omega\nabla_{\nu}\phi-\frac{1}{3}\nabla_{\rho}\Omega\nabla^{\rho}\phi\right)\nonumber
\\&\qquad\qquad\qquad+\frac{\tilde{\omega}-\omega}{\phi^{2}}\left(h^{\mu\nu}\nabla_{\mu}\phi\nabla_{\nu}\phi
-\frac{1}{3}\nabla_{\rho}\phi\nabla^{\rho}\phi\right)+\frac{\Omega^{4}\tilde{V}-V}{3\phi}\Bigg].
\end{align}
We see here the appearance of two new integrals beside the initial integral giving the Hawking-Hayward mass in (\ref{HHinST}). In analogy with what we argued about the transformation of the purely geometric definition, the two last integrals in (\ref{ConfHHinST}) could be interpreted as coming from the 'induced Hamiltonian' by the conformal transformation as well as the 'interaction Hamiltonian' of $\phi$ with the conformal geometry, both projected on the 2-surface $\mathcal{S}$.

Let us now conduct an analysis on the result (\ref{ConfHHinST}) analogous to what we did for the result (\ref{ConfHHinGR}) found within general relativity. Let us start with the simple case of a conformal transformation of the form $\Omega(x)=\phi(x)$. For this simple special case we know \cite{FaraoniBook} that the action, as well as the field equations, of the Brans-Dicke theory remain invariant, provided that the potential of the scalar field $\phi$ transforms like $\tilde{V}(\tilde{\phi})=\Omega^{-4}V(\phi)$ and that the Brans-Dicke parameter remains invariant; that is $\tilde{\omega}(\tilde{\phi})=\omega(\phi)$. Substituting these three equalities inside (\ref{ConfHHinST}), one finds that the latter reduces to,
\begin{align}\label{ConfHHinSTOmega=phi}
\tilde{m}_{_{HH}}&=\sqrt{\tilde{A}}\int_{\mathcal{S}}\frac{\mu\phi}{\Omega^{2}}\Bigg[h^{\mu\rho}h^{\nu\sigma}C_{\mu\nu\rho\sigma}-2\omega_{\mu}\omega^{\mu}+\frac{8\pi}{\phi}\left(h^{\mu\nu}T_{\mu\nu}
-\frac{2T}{3}\right)\nonumber
\\&\qquad\qquad\qquad+\frac{h^{\mu\nu}\nabla_{\mu}\nabla_{\nu}\phi}{\phi}+\frac{\omega}{\phi^{2}}\left(h^{\mu\nu}\nabla_{\mu}\phi\nabla_{\nu}\phi
-\frac{1}{3}\nabla^{\rho}\phi\nabla_{\rho}\phi\right)+\frac{V}{3\phi}\Bigg]\nonumber
\\&+2\sqrt{\tilde{A}}\int_{\mathcal{S}}\frac{\mu\phi}{\Omega^{2}}\Bigg[h^{\mu\nu}
\left(\frac{2\nabla_{\mu}\Omega\nabla_{\nu}\Omega}{\Omega^{2}}
-\frac{\nabla_{\mu}\nabla_{\nu}\Omega}{\Omega}\right)-\frac{\nabla_{\rho}\Omega\nabla^{\rho}\Omega}{\Omega^{2}}\Bigg].
\end{align}
It is already clear that this result is again different from (\ref{ConfGeoHH}), which has been obtained from the purely geometric form (\ref{HHReducedDefinition}) of the Hawking-Hayward mass. That is, although one might have argued that one finds in general relativity different results whether one uses the field equations or not because general relativity is not conformally invariant, and that the issue might be settled by going to the conformally invariant scalar-tensor theories, the present result shows that the conflict between the two approaches does not disappear.

To make things even more transparent, let us examine the case of spherical symmetry for which $\Omega$ could be taken out of the integrals and the equality $\tilde{A}=\Omega^{2}A$ could be used. The result (\ref{ConfHHinSTOmega=phi}) then simply reads,
\begin{align}\label{ConfHHinSTOmega=phiSphere}
\tilde{m}_{_{HH}}=\Omega^{-1}\left(m_{_{HH}}+\phi\,m_{HH}^{I}\right).
\end{align}
Here we have again written the different contributions separately. The first is just the initial Hawking-Hayward mass containing both the contribution of ordinary matter, via its energy-momentum tensor $T_{\mu\nu}$, and the contribution of pure geometry, including $\phi$. The second term is the contribution of the induced geometric energy through $\Omega$.

We clearly recognize here the familiar behavior of real masses under conformal transformation, i.e. the overall $\Omega^{-1}$ factor. Thus, unlike in general relativity, in conformally invariant theories there is no distinction between material and geometric entities as the whole mass transforms as normal masses do. It must be noted here that this result didn't depend on the nature of the material Lagrangian $L_{matter}$ appearing inside the action. In fact, it is well-known (see e.g. \cite{Quiros}) that the field equations of scalar-tensor theories, including matter, are invariant only when the matter Lagrangian depends also on the scalar field $\phi$, that is $L_{mattar}=L_{matter}(g,\phi)$ and a non-minimal coupling of matter with $\phi$ exists. No such constraint is required here as no condition is imposed on the energy-momentum tensor appearing inside the integral (\ref{HHinST}).

Let us now examine the case of the more general conformal transformations of the form $\Omega=\phi^{\alpha}$. We know \cite{FaraoniBook} that for that case one has invariant action as well as field equations only for a specific behavior of the Brans-Dicke parameter $\omega(\phi)$ under conformal transformations. Let us examine here how, and which, conditions are imposed by the Hawking-Hayward mass. Substituting $\Omega=\phi^{\alpha}$ in (\ref{ConfHHinST}), we find
\begin{align}\label{ConfHHinSTOmega=phialpha}
\tilde{m}_{_{HH}}&=\sqrt{\tilde{A}}\int_{\mathcal{S}}\frac{\mu\phi}{\Omega^{2}}\Bigg[h^{\mu\rho}h^{\nu\sigma}C_{\mu\nu\rho\sigma}-2\omega_{\mu}\omega^{\mu}+\frac{8\pi}{\phi}\left(h^{\mu\nu}T_{\mu\nu}
-\frac{2T}{3}\right)\nonumber
\\&\qquad\qquad\qquad+\frac{h^{\mu\nu}\nabla_{\mu}\nabla_{\nu}\phi}{\phi}+\frac{\omega}{\phi^{2}}\left(h^{\mu\nu}\nabla_{\mu}\phi\nabla_{\nu}\phi
-\frac{1}{3}\nabla^{\rho}\phi\nabla_{\rho}\phi\right)+\frac{V}{3\phi}\Bigg]\nonumber
\\&+2\sqrt{\tilde{A}}\int_{\mathcal{S}}\frac{\mu\phi}{\Omega^{2}}\Bigg[h^{\mu\nu}\left(\frac{5\alpha-3}{\alpha}\frac{\nabla_{\mu}\Omega\nabla_{\nu}\Omega}{\Omega^{2}}
-\frac{\nabla_{\mu}\nabla_{\nu}\Omega}{\Omega}\right)+\frac{1-2\alpha}{\alpha}\frac{\nabla_{\rho}\Omega\nabla^{\rho}\Omega}{\Omega^{2}}\Bigg]\nonumber
\\&+2\sqrt{\tilde{A}}\int_{\mathcal{S}}\frac{\mu\phi}{\Omega^{2}}\left[\frac{(1-2\alpha)^2\tilde{\omega}-\omega}{2\alpha^{2}}
\left(h^{\mu\nu}\frac{\nabla_{\mu}\Omega\nabla_{\nu}\Omega}{\Omega^{2}}-\frac{1}{3}\frac{\nabla_{\rho}\Omega\nabla^{\rho}\Omega}{\Omega^{2}}\right)\right].
\end{align}

First, we notice that whatever value of $\alpha$ one uses, this expression could never be made completely identical to the expression (\ref{ConfGeoHH}) obtained from the purely geometric expression because of the presence of $1/\Omega^{2}$ inside every integral in (\ref{ConfHHinSTOmega=phialpha}). Thus, no special choice of the conformal transformation of $\omega$ could make the Hawking-Hayward mass 'per unit length' $m_{_{HH}}/\sqrt{A}$ invariant anymore, and hence transform as a length, as was the case in (\ref{ConfGeoHH}) found from the geometric approach.

Now, the result (\ref{ConfHHinSTOmega=phialpha}) could still be made to resemble formally the expression (\ref{ConfHHinSTOmega=phi}), in the sense that it reduces to the original mass, weighed by a certain factor, plus a term coming from the energy induced by the work due to the deformation $\Omega$. More precisely, we demand that the two last integrals in (\ref{ConfHHinSTOmega=phialpha}), representing the geometrically induced energy, should sum up and give the same form as the one appearing both in the second integral in (\ref{ConfHHinSTOmega=phi}) and in (\ref{mHHInduced}) coming from the purely geometric definition (\ref{HHReducedDefinition}) of the mass. We easily find that this requirement is satisfied if and only if (with $\alpha\neq1/2$),
\begin{equation}\label{omegaConstraint}
\tilde{\omega}=\frac{\omega-6\alpha(\alpha-1)}{(1-2\alpha)^{2}},
\end{equation}
in which case (\ref{ConfHHinSTOmega=phialpha}) reduces to (\ref{ConfHHinSTOmega=phi}). Now the condition (\ref{omegaConstraint}) is nothing but the exact condition that makes the Brans-Dicke theory conformally invariant \cite{FaraoniBook,Faraoni1,Faraoni2}. This result is actually hardly surprising, given that we found it by demanding invariance within Brans-Dicke theory. Still, this result could not have been guessed \textit{a priori} neither, given that what has been substituted inside the mass integral (\ref{HHinST}) in terms of matter was not the Einstein tensor $R_{\mu\nu}-\frac{1}{2}g_{\mu\nu}R$, living in the left-hand side of the field equations, but rather the combination $R_{\mu\nu}-\frac{1}{6}g_{\mu\nu}R$ which is not relevant for the field equations. Therefore, it is not obvious that asking for the invariance of the field equations would also guarantee the invariance of the mass.

In summary then, we learn that when it comes to conformal transformations the purely geometric definition of the Hawking-Hayward mass (\ref{HHDefiniton}) is fundamentally different from the energy content (\ref{HHinGR}) (within general relativity) or (\ref{HHinST}) (within scalar-tensor theories) one extracts from this mass through the field equations, and this even within conformally invariant scalar-tensor theories of gravity.

\section{Similarities between Hawking-Hayward and Misner-Sharp}\label{sec:5}
In this section we are going to discuss the similarities between the conformal transformations of the two quasi-local mass concepts. In fact, although it is well-known that the two definitions coincide in spherical symmetry \cite{Hay}, nothing guarantees that they would also display the same conformal behavior after imposing the field equations on the Hawking-Hayward mass. Let us start by making a comparison between the conformal transformations of the two masses when both are based on their geometric definition.

The first similarity point is that both masses display the same 'wrong' overall factor under conformal transformations when used in their purely geometric definitions. For the Hawking-Hayward mass this behavior is clearly displayed for the spherically symmetric case in (\ref{ConfGeoHHSphere}). For the Misner-Sharp mass $m_{_{MS}}$, whose geometric definition is $m_{_{MS}}=R(1-\partial_{\mu}R\partial^{\mu}R)/2G$ \cite{MS} for a spherical region of areal radius $R(t,r)$, the same behavior is displayed in equations (2.10) and (2.13) of Ref.~\cite{Hammad}, for the Friedmann-Lema\^{\i}tre-Robertson-Walker (FLRW) geometry and the Schwarzschild geometry, respectively.

The second similarity consists of the fact that for the conformally invariant Brans-Dicke scalar-tensor theory both masses display the 'right' conformal factor. For the Hawking-Hayward mass, this behavior shows up clearly for the spherically symmetric case in (\ref{ConfHHinSTOmega=phiSphere}). For the Misner-Sharp mass, the same behavior is displayed in equation (2.19) of Ref.~\cite{Hammad}, for an FLRW geometry.

Let us now make a comparison between the two masses when the Hawking-Hayward mass is expressed in terms of matter through the field equations and the Misner-Sharp mass is expressed in terms of the density of its matter content in the form \cite{MS}:
\begin{equation}\label{MaterialMS}
m(t,r)=\int4\pi R^{2}\rho\frac{\partial R}{\partial r}\mathrm{d}r.
\end{equation}
The result found for the Hawking-Hayward mass in the spherically symmetric case is displayed in (\ref{ConfHHinGRSphere}) and the result found for the Misner-Sharp mass is displayed in equation (3.13) of Ref.~\cite{Hammad}, which we recall here:
\begin{equation}\label{MSResult1}
\tilde{m}(t,r)=\frac{m(t,r)}{\Omega}+\int\left[m(t,r)+4\pi R^{3}\rho\right]\frac{\Omega'}{\Omega^{2}}\mathrm{d}r.
\end{equation}

We see that just as it happens for the Misner-Sharp mass in (\ref{MSResult1}), the Hawking-Hayward mass in (\ref{ConfHHinGRSphere}) splits into two, the purely material content, which transforms as usual masses, and an additional term. The difference, though, is that the additional term in the case of Misner-Sharp has also the right conformal factor, albeit inside an integral, whereas the additional term in (\ref{ConfHHinGRSphere}) comes with the wrong factor $\Omega$. The reason is that the additional term in (\ref{ConfHHinGRSphere}) is purely geometric in contrast to the additional integral in (\ref{MSResult1}), which was interpreted in Ref.~\cite{Hammad} as being due to the interaction energy between matter and the conformal geometry. It is as if basing the Msiner-Sharp mass on its material content makes the mass miss the independent behavior of the purely geometric energy not missed by the Hawking-Hayward mass. By writing the latter with the help of the field equations as in (\ref{HHinGR}), we merely separated the material content from what is left as pure geometry inside the mass. This was not the case, however, with the Misner-Sharp mass, where one tracks down only the material content when using the definition (\ref{MaterialMS}).

The more subtle difference, however, is that the Hawking-Hayward mass displays an induced mass after a conformal transformation within scalar-tensor theories whenever it is extracted from the field equations of the theory. This fact appears clearly in (\ref{ConfHHinSTOmega=phiSphere}), where we have denoted the induced term by $m_{_{HH}}^{I}$. Such a term did not appear neither in the Misner-Sharp case nor even in the Hawking-Hayward case within general relativity. The absence of such a term in both is actually not a coincidence. On one hand, its absence in the Misner-Sharp case is due to the fact that by basing the latter only on its material content one simply discards pure geometry from which the term arises. On the other hand, its absence in the Hawking-Hayward case within general relativity is due to the fact that when using the field equations, the part of geometry that would have given birth to the induced energy, if it were not transformed into matter, is the combination $R_{\mu\nu}-\frac{1}{6}g_{\mu\nu}R$ which gets replaced by $8\pi(T_{\mu\nu}-\frac{1}{3}g_{\mu\nu}T)$ via Einstein field equations. In scalar-tensor theories the problem is automatically resolved as pure geometry effects are saved, even by using the field equations, thanks to the presence of the field $\phi$ which restores them via its energy-momentum tensor.

It is actually possible to also recover within general relativity the exact induced energy contribution in both masses after a conformal transformation provided that one carefully includes its contribution inside the integrals. We shall show this, first for the Hawking-Hayward mass, and then we expose the procedure for the Misner-Sharp case\footnote{This procedure has not been exposed in Ref.~\cite{Hammad}.}.

For the case of the Hawking-Hayward mass, one needs just to extract from the term converted into matter the induced energy it gives birth to. Indeed, since we have the following transformation:
\begin{equation}\label{OrigineHHInduced}
\tilde{R}_{\mu\nu}-\frac{1}{6}\tilde{g}_{\mu\nu}\tilde{R}=R_{\mu\nu}-\frac{1}{6}g_{\mu\nu}R+\frac{4\nabla_{\mu}\Omega\nabla_{\nu}\Omega}{\Omega^{2}}
-\frac{2\nabla_{\mu}\nabla_{\nu}\Omega}{\Omega}-g_{\mu\nu}\frac{\nabla_{\rho}\Omega\nabla^{\rho}\Omega}{\Omega^{2}},
\end{equation}
multiplying the sum of the three additional terms in (\ref{OrigineHHInduced}) by $\mu h^{\mu\nu}$ and then integrating it over the 2-surface $\mathcal{S}$ one finds exactly, after multiplying the result by $\sqrt{A}$, the missing induced energy (\ref{mHHInduced}) that didn't rise within general relativity.

For the case of the Misner-Sharp mass, one needs just to extract the induced energy from the right-hand side of the transformed field equations $\tilde{G}_{\mu}^{\;\nu}=\kappa\Omega^{-2}T_{\mu}^{\;\nu}+\Omega^{-2}T_{\mu}^{\Omega\nu}$, where $T_{\mu\nu}^{\Omega}$ is the induced energy-momentum tensor \cite{Dabrowski}. All one has to do then is to write (\ref{MaterialMS}) in the Einstein frame and use instead of $\rho$ the induced density $\rho^{I}$ to be extracted from the $tt$-component of the induced energy-momentum tensor,
\begin{equation}\label{InducedT}
T_{\mu\nu}^{\Omega}=\frac{4\nabla_{\mu}\Omega\nabla_{\nu}\Omega}{\Omega^{2}}-\frac{2\nabla_{\mu}\nabla_{\nu}\Omega}{\Omega}
+g_{\mu\nu}\left(2\frac{\square\Omega}{\Omega}-\frac{\nabla_{\rho}\Omega\nabla^{\rho}\Omega}{\Omega^{2}}\right).
\end{equation}
For the case of FLRW geometry, the metric is $\mathrm{d}s^{2}=-\mathrm{d}t^{2}+a^{2}(t)(\mathrm{d}r^{2}+r^{2}\mathrm{d}o^{2})$, where $a(t)$ is the positive scale factor and $\mathrm{d}o^{2}=\mathrm{d}\theta^{2}+\sin\theta^{2}\mathrm{d}\phi^{2}$ is the metric on the unit-sphere. One then has the simple conformal transformation $\tilde{R}(t,r)=\Omega(t)R(t,r)$ of the areal radius, since the latter is $R(t,r)=a(t)r$ and, to preserve the homogeneity of the FLRW Universe, the conformal factor $\Omega$ should not depend on $r$. The induced-energy density in this geometry is given by $\rho^{I}=-\frac{1}{\kappa}\Omega^{-2}T_{t}^{\Omega t}$. The computation of the integral is then made easy and one finds\footnote{We restore here the gravitational constant $G$ for ease of comparison with the results of Ref.~\cite{Hammad}},
\begin{equation}\label{OrigineMSInduced1}
\tilde{m}_{_{MS}}^{I}=\int4\pi\tilde{R}^{2}\rho^{I}\frac{\partial\tilde{R}}{\partial r}\mathrm{d}r
=\frac{\Omega R^{3}}{2G}\left(2H\frac{\dot{\Omega}}{\Omega}+\frac{\dot{\Omega}^{2}}{\Omega^{2}}\right).
\end{equation}
This is exactly the induced term found in equation (2.10) of Ref.~\cite{Hammad} for the conformal transformation of the Misner-Sharp mass when based on its purely geometric definition.

For the case of the Schwarzschild geometry whose black hole's mass is $M$ the metric reads, $\mathrm{d}s^{2}=-f(r)\mathrm{d}t^{2}+f^{-1}(r)\mathrm{d}r^{2}+r^{2}\mathrm{d}o^{2}$, where $f(r)=1-2GM/r$. One has the conformal transformation of the areal radius, $\tilde{R}(t,r)=\Omega(t,r)R(r)$, but now the conformal factor depends both on time $t$ and the coordinate $r$. The areal radius being given by $R(r)=r$. Using this metric, one easily computes the induced energy density $\frac{-1}{\kappa}\Omega^{-2}T_{t}^{\Omega t}$ to be,
\begin{equation}\label{InducedrhoSch}
\rho^{I}=\frac{1}{\kappa}\left(\frac{3\dot{\Omega}}{f\Omega^{4}}-\frac{f'\Omega'}{\Omega^{3}}
-\frac{2\Omega''f}{\Omega^{3}}-\frac{4\Omega'f}{r\Omega^{3}}+\frac{f\Omega'^{2}}{\Omega^{4}}\right).
\end{equation}
In contrast to the previous case, however, the energy-flux $T_{t}^{\Omega r}$ is also not zero and, hence, just integrating the induced energy density $\rho^{I}=-\frac{1}{\kappa}\Omega^{-2}T_{t}^{\Omega t}$ would not give exactly the same induced energy found in equation (2.13) of Ref.~\cite{Hammad} from the purely geometric definition of the Misner-Sharp mass. In fact, using (\ref{InducedrhoSch}), we find after some integrations by parts,
\begin{equation}\label{OrigineMSInduced2}
\tilde{m}_{_{MS}}^{I}=\int4\pi\tilde{R}^{2}\rho^{I}\frac{\partial\tilde{R}}{\partial r}\mathrm{d}r
=\frac{\Omega r^{3}}{2G}\left(\frac{\dot{\Omega}^{2}}{\Omega^{2}f}-2f\left[\frac{\Omega'}{r\Omega}
+\frac{\Omega'^{2}}{2\Omega^{2}}\right]\right)+...,
\end{equation}
where the ellipsis represent the additional integrals yet to be performed. The terms displayed here constitute all that was found in equation (2.13) of Ref.~\cite{Hammad}. Thus, unlike for the homogeneous FLRW geometry, the purely material integral (\ref{MaterialMS}) does not give, when using $\rho^{I}$, the same result for induced energy with a general conformal factor $\Omega$. This can be traced back to the fact that the integral definition of the Misner-Sharp mass (\ref{MaterialMS}) was obtained from Einstein equations \cite{MS}. For a conformal factor that depends only on time, the conformally transformed Einstein equations still yield the same integral relation (\ref{MaterialMS}) with tildes over the letters in the conformal frame \cite{Hammad}. For a more complicated conformal factor $\Omega(t,r)$, though, the field equations do not yield simply the right-hand side of (\ref{MaterialMS}) in the Einstein frame but the result looks like this,
\begin{equation}\label{MSCompleteMaterial}
\tilde{m}(t,r)=\int4\pi\tilde{R}^{2}\tilde{\rho}\frac{\partial\tilde{R}}{\partial r}\mathrm{d}r+...,
\end{equation}
where the ellipsis again represent both the induced geometric energy and interaction terms between matter and geometry. Adopting only the displayed terms in (\ref{MSCompleteMaterial}), as it was done in Ref.~\cite{Hammad}, will then yield the correct partial answer for the matter-geometry coupling but will automatically miss the induced geometric energy.

Finally, although beyond the scope of the present paper, it should be noted here that it is still possible to recover the complete induced geometric energy together with all the possible couplings of the latter with matter in a covariant way. In fact, one needs only to use the conservation equation through the Bianchi identity in the Einstein frame. One has then the conservation of the total energy-momentum tensor, $\tilde{\nabla}_{\nu}\tilde{T}_{\mathrm{total}}^{\mu\nu}=\tilde{\nabla}_{\nu}\left[\Omega^{-4}\left(\kappa T^{\mu\nu}+T_{\Omega}^{\mu\nu}\right)\right]=0$ \cite{Dabrowski}, and this equation takes into account not only the induced energy, but also the interactions between matter and geometry.

\section{conclusion}
We have examined in this paper the behavior of the Hawking-Hayward quasi-local mass under conformal transformations of spacetime. We have viewed the transformations from different angles and looked at its different facets. We first recalled the results found recently in the literature about the transformation of this mass whenever one relies entirely on the purely geometric definition of the latter. In the simple case of spherical symmetry the resulting transformed mass (\ref{ConfGeoHHSphere}) could be separated into two parts, the initial mass and a new part which we interpreted as coming from the geometrically induced energy. An overall factor was present, though, in contrast to what usual masses exhibit under a conformal transformation. We explained the 'strange' multiplicative factor $\Omega$ that appears as due to the fact that what the formula represents geometrically is not the mass/energy itself but the product of $G$ times mass/energy, and hence has automatically the dimensions of length.

The Hawking-Hayward mass recovers the true conformal behavior of real masses whenever one uses the field equations of the gravitational theory and substitutes the material equivalent of the different geometric entities inside the integral defining the mass. In so doing, we saw that in general relativity the material part in (\ref{ConfHHinGRSphere}) behaves as usual masses, by being multiplied by $\Omega^{-1}$, but the geometric part still exhibits the 'wrong' factor $\Omega$. By going to conformally invariant theories of gravity, though, we satisfactorily recovered in (\ref{ConfHHinSTOmega=phiSphere}) the right overall multiplicative factor $\Omega^{-1}$, shared by the material, the geometric, and the geometrically induced part. Thus, while the right factor $\Omega^{-1}$ started to manifest itself in general relativity only on the matter part, when going to conformally invariant theories the whole quasi-local mass acquires the right factor. In contrast to scalar-tensor theories, however, there is no geometrically induced part within general relativity. The reason being that by using the field equations of general relativity one entirely converts into matter what would have given rise to the geometrically induced energy. As we saw, though, it is possible to recover the induced energy by hand if one extracts it from the geometry that gets converted into matter. In Sec.~\ref{sec:5} we saw how similar intricacies manifest themselves in the case of the Misner-Sharp mass under conformal transformations.

In Sec.~\ref{sec:4}, we saw that the quasi-local mass allows one to recover the familiar condition (\ref{omegaConstraint}) imposed on the transformation of the Brans-Dicke parameter $\omega(\phi)$ to insure conformal invariance of Brans-Dicke theory. It is remarkable that one recovers the same constraint without asking for the invariance of the field equations in the first place. All that was needed was that one asks for the invariance of the form of the geometrically induced mass. Of course, the field equations are certainly behind this since they have been used to convert geometry into matter. Nevertheless, the result is still not obvious at all given that we were not trying to cancel the induced part, as is the case when seeking invariance of the field equations, but on the contrary to keep that additional part similar in form to the one rising from the purely geometric definition of the mass or from the simple invariant theory whose parameter transforms as $\tilde{\omega}=\omega$ for the conformal factor $\Omega=\phi$. It is therefore not surprising that the condition on $\omega$ recovered didn't depend in any way on the precise coupling of matter with the field $\phi$ as is the case when seeking invariance of the field equations.

For the sake of completeness, we should notice here that we focused in this paper mainly on scalar-tensor theories of gravity although the case of $f(R)$ gravity is not less important. One reason is that it is well-known that $f(R)$-gravity is just a subclass of scalar-tensor theories of gravity, as a redefinition of the variables inside the $f(R)$-gravity action turns the latter into a Brans-Dicke action with vanishing Brans-Dicke parameter $\omega$ but a non-zero potential $V(\phi)$ for the induced scalar field $\phi$ (see \textit{e}.\textit{g}. \cite{Capozziello}). The other reason is that the main purpose of this paper was not to study how the Hawking-Hayward mass behaves within other theories of gravity, but only to bring into light the peculiarities in its behavior under conformal transformations whenever one relies on a specific theory to define it. For that purpose we used the simplest prototype of extended theories of gravity, namely, Brans-Dicke theory, although other theories might as well be used.

Finally, given the importance of both the quasi-local mass concept and the conformal transformations in the study of spacetime thermodynamics, non-trivial facts about the latter could be learned from the results obtained here. Given that black hole thermodynamics relies heavily on the horizon geometry to extract temperature and entropy (see \textit{e}.\textit{g}. the review \cite{Nielsen}), converting geometry into its matter equivalent might not give the correct results. However, the possibility of separating matter from geometry without destroying completely the effects of the latter within the Hawking-Hayward mass concept, might still give sensible results. A future study of this issue will be done in a forthcoming work.
\\\\
\textbf{Acknowledgement}:
I am grateful to Valerio Faraoni for the helpful discussion.

\end{document}